\documentclass[prb,superscriptaddress, twocolumn]{revtex4-2} 
\usepackage{xcolor}
\usepackage{hyperref}
\usepackage{amsmath}
\usepackage{amsfonts}
\usepackage{natbib}
\usepackage{graphicx}
\usepackage{graphicx, nicefrac}
\usepackage{dcolumn}
\usepackage{float}
\usepackage{subfig}
\usepackage{bm}
\usepackage{xfrac} 

\usepackage{braket} 
\usepackage{svg}
\usepackage{float}

\usepackage{soul}

\begin{document}

\title{Dark resonance spectra of trapped ions under the influence of micromotion}
\author{Nicol\'as A. Nu\~nez Barreto}
\email{nnunez@df.uba.ar}
\affiliation{Universidad de Buenos Aires, Facultad de Ciencias Exactas y Naturales, Departamento de F\'isica,
Laboratorio de Iones y \'Atomos Fr\'ios, Pabell\'on 1, Ciudad Universitaria, 1428 Buenos Aires, Argentina}
\affiliation{CONICET - Universidad de Buenos Aires, Instituto de F\'isica de Buenos Aires (IFIBA), Pabell\'on 1,
Ciudad Universitaria, 1428 Buenos Aires, Argentina}
\author{Muriel Bonetto}
\affiliation{Universidad de Buenos Aires, Facultad de Ciencias Exactas y Naturales, Departamento de F\'isica,
Laboratorio de Iones y \'Atomos Fr\'ios, Pabell\'on 1, Ciudad Universitaria, 1428 Buenos Aires, Argentina}
\affiliation{CONICET - Universidad de Buenos Aires, Instituto de F\'isica de Buenos Aires (IFIBA), Pabell\'on 1,
Ciudad Universitaria, 1428 Buenos Aires, Argentina}
\author{Marcelo A. Luda}
\affiliation{Universidad de Buenos Aires, Facultad de Ciencias Exactas y Naturales, Departamento de F\'isica,
Laboratorio de Iones y \'Atomos Fr\'ios, Pabell\'on 1, Ciudad Universitaria, 1428 Buenos Aires, Argentina}
\affiliation{CEILAP, CITEDEF, Buenos Aires, 1603, Argentina}
\author{Cecilia Cormick}
\affiliation{Instituto de F\'isica Enrique Gaviola, CONICET and Universidad Nacional de C\'ordoba,
Ciudad Universitaria, X5016LAE, C\'ordoba, Argentina}
\author{Christian T. Schmiegelow}
\affiliation{Universidad de Buenos Aires, Facultad de Ciencias Exactas y Naturales, Departamento de F\'isica,
Laboratorio de Iones y \'Atomos Fr\'ios, Pabell\'on 1, Ciudad Universitaria, 1428 Buenos Aires, Argentina}
\affiliation{CONICET - Universidad de Buenos Aires, Instituto de F\'isica de Buenos Aires (IFIBA), Pabell\'on 1,
Ciudad Universitaria, 1428 Buenos Aires, Argentina}

\date{\today}

\begin{abstract}

We study the influence of micromotion on the spectrum of trapped ions with a lambda-type level scheme, leading to dark resonances due to coherent population trapping. We work with calcium ions trapped in a ring-shaped Paul trap, in which one can compensate excess micromotion for only one ion of the crystal. 
We observe that micromotion affects
the shapes of the dark resonances and causes the appearance of ``echoes'' separated by intervals given by the drive frequency. 
We present a theoretical model that provides good fits to the measurements and can be used to estimate the amplitude of the micromotion modulation of the atomic motion. 
We estimate an effective temperature of the ions from the spectra and observe clear micromotion heating as well as impaired cooling for sufficiently large excess micromotion.

\vspace{5mm}

\end{abstract}

\maketitle

\section{Introduction}

Trapped-ion systems are versatile platforms which in the last decades led to the development of quantum technologies such as quantum simulators~\cite{porras2004effective,friedenauer2008simulating}, optical clocks~\cite{rosenband2008frequency,chwalla2009absolute,chou2010frequency,huntemann2012high}, and small-scale prototype quantum computers~\cite{cirac1995quantum,blatt2008entangled}. They have also served as a platform for the study of fundamental physics at the quantum level and for the search of new physics via precision measurements~\cite{safronova2018search,shaniv2018new,berengut2018probing,keller2019probing,counts2020evidence}. 

Paul traps, and more generally radiofrequency (RF) traps, employ electric RF fields to confine ions in vacuum~\cite{neuhauser1980localized,paul1990electromagnetic}. 
The dynamics of the ions within these traps can generally be understood in terms of two contributions: macro or secular motion, and micromotion. For appropriate trap parameters, the oscillating RF field creates an effective static harmonic potential with a characteristic frequency that is much slower than that of the RF field; the movement of the ions due to this effective potential is called secular motion. This motion can be externally controlled by cooling, coherent excitation and preparation of quantum motional states~\cite{leibfried2003quantum}. On the other hand, the oscillating character of the trapping field creates a driven motion at the RF drive frequency, named micromotion. The micromotion amplitude is generally small and, in lowest-order approximation, proportional to the displacement of the ions with respect to the center of the trap, where the RF field vanishes. 

As has been shown in the last years, it is possible to use micromotion as a resource to enhance the speed and fidelity of entangling gates in trapped ions~\cite{bermudez2017micromotion,ratcliffe2020micromotion,gaebler2021suppression}. Nevertheless, it remains generally desirable to have low micromotion, for example to achieve efficient ion cooling~\cite{devoe1989role,cirac1994laser,sikorsky2017doppler}, reduce heating~\cite{turchette2000heating,brouard2001heating}, and reduce Doppler related shifts in trapped ion clocks~\cite{keller2015precise,ludlow2015optical}. With this goals in mind, several techniques were developed that allow for the reduction of micromotion, usually referred to as micromotion compensation~\cite{berkeland1998minimization,ibaraki2011detection,higgins2021micromotion}. 

Micromotion-related spurious effects are particularly problematic in ion crystals. 
In this case, the micromotion will affect the ions differently depending on their locations within the trap~\cite{van2022rf}. 
The geometry of the trap and the potentials used then determine the possible ion crystal shapes and the properties of the micromotion for each ion. For instance, the linear Paul trap is specially well-suited to work with linear ion chains since ions can be made to lie along an axis of vanishing oscillating field. On the other hand, in three-dimensional (3D) traps like ring traps the RF field vanishes at only one point in space, so excess micromotion can at best be fully compensated for a single ion.

Micromotion has important modulation effects in the atomic spectra of ions, which are different depending on the width of the relevant spectral features. Considering that the typical drive frequency in RF traps is usually in the MHz regime, micromotion will be reflected either as broadening of the spectral lines, or echoes at the drive frequency, depending on whether the width of the spectral lines is wider or narrower than the drive frequency, respectively. The first case is typical of dipole transitions, and its effects are commonly used to compensate excess micromotion. The procedure consists in driving a fluorescence transition with a detuning equal to the drive frequency and minimizing the fluorescence when varying the voltage of several electrodes. This reduces the micromotion, which in turn sharpens the spectrum~\cite{pruttivarasin2014direct}. On the other hand, for narrow transitions like quadrupolar or stimulated Raman  transitions, the spectral lines are replicated as smaller echoes separated by the drive frequency~\cite{peik1999sideband,goham2022resolved}.

A mixed behavior between these two regimes can be found in dark resonance spectra, which we study in detail in this article. The atomic spectra of atoms with lambda-type level schemes present narrow negative peaks, known as dark resonances, when the detunings of the lasers that drive the transitions from the lower states to the excited one are similar. This is due to coherent population trapping (CPT) phenomena~\cite{arimondo1976nonabsorbing,gray1978coherent,vanier2005atomic}. The widths of the dark resonances are typically much smaller than those of dipolar transitions. 

\begin{figure}[h!]
\centering
\includegraphics[width=0.5\textwidth]{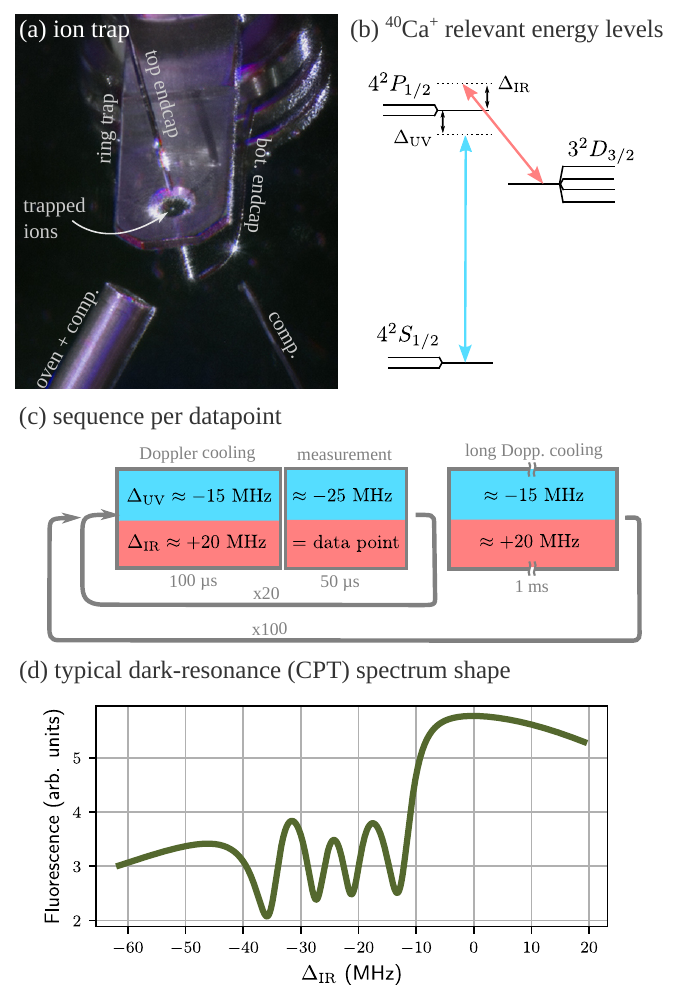}
\caption[caption of figure1]{\label{fig:fig1} (a) Photo of the ring trap\footnote{Photo taken by Franco Meconi, \url{https://www.instagram.com/terrazaalcosmos/ }}, where the electrodes are indicated. At the trap center, fluorescence from trapped ions can be seen. (b) Relevant level scheme for the $^{40}$Ca$^+$ ion with the two lasers, labelled UV and IR, that excite each of the transitions. (c) Pulsed sequence for measuring dark resonance spectra. An initial Doppler cooling of 100~$\mu$s is followed by a measurement of 50~$\mu$s. After 20 repeats, a long Doppler cooling of 1~ms is performed to avoid excess heating. The whole sequence is repeated 100 times to increase the collected ion light and thus reduce statistical uncertainties. (d) A typical dark resonance spectrum with four dark resonances that emerge when the polarization of both lasers is $\sigma_+ + \sigma_-$, i.e., linear and perpendicular to the magnetic field that is used to obtain a Zeeman splitting of the sublevels in each manifold.}
\end{figure}

For multi-level systems, many dark resonances can emerge depending on the polarizations of the lasers and the geometry of the optics. A typical dark resonance spectrum is shown in Fig.~\ref{fig:fig1}(d), where the polarizations are such that four dark resonances emerge. The widths are related to the product of the two Rabi frequencies and the dephasing mechanisms at work, and reach values of up to a few MHz. For these spectra, the micromotion will manifest as a mixture of further broadening of the already broad main line together with echoes of the dark resonances which appear as dips within the main line. 

In this paper we present a detailed study of the effects of micromotion in dark resonance spectra in trapped ions. We work with $^{40}$Ca$^+$ ions confined in a 3D ring-shaped Paul trap. 
By displacing a single ion within the trap we measure its dark resonance spectrum under different micromotion conditions. We show how to retrieve the modulation factor of the ion using a theoretical model to fit the spectra. We also estimate an effective temperature of the ion, evidencing the presence of an additional heating mechanism due to micromotion known as RF heating~\cite{blumel1989chaos,kalincev2021motional}. Moreover, we measure the spectra of more than one ion at the same time and retrieve the modulation factor of each ion through a fit to the model for the full measurement.

The paper is organized as follows. In Sec.~\ref{sec:sec2} we present the experimental setup and the measurement algorithm as well as the compensation scheme. In Sec.~\ref{sec:sec3} we show how to calculate dark resonance spectra with and without micromotion. In Sec.~\ref{sec:sec4} we present results for micromotion-modulated spectra with a single ion and analyze how it alters the final temperature of the ion. Finally, in Sec.~\ref{sec:sec5} we present results on dark resonance spectra for two ions and study how to model the sum of their spectra.

\section{Experimental setup and compensation scheme}
\label{sec:sec2}

We perform our experiments with singly-ionized $^{40}$Ca atoms trapped in a 3D ring-shaped Paul trap, as shown in Fig.~\ref{fig:fig1}(a). The ions sit at the center of the ring-shaped electrode, set to a voltage of $\sim~300~$V$_{\text{pp}}$ oscillating at 22.1~MHz which is responsible for the radial trapping. The axial confinement, on the other hand, is produced by two endcaps, top and bottom, with $\sim -0.5~$V each. One extra electrode located below the trap is used to compensate the position of the ion. The oven, which is the source of neutral $^{40}$Ca atoms, is also used as an additional electrode for compensation. In this work, we will vary the voltage of the top endcap electrode to controllably add micromotion to the ions.

The calcium ion of interest has a multi-lambda type electronic level system, schematized in Fig.~\ref{fig:fig1}(b). The two ground 4$^2$S$_{1/2}$ states are connected through a dipole transition near 397$~$nm (which we label UV for ``ultraviolet'') to two excited 4$^2$P$_{1/2}$ states, with a lifetime of~$\sim 7$~ns~\cite{hettrich2015measurement}. From this P level, the ion has a probability of 0.94 to decay back to the S states, and a probability of 0.06 to decay to one of the four metastable 3$^2$D$_{3/2}$ states~\cite{ramm2013precision,barreto2023transient}. These metastable states are connected to the excited ones through another dipolar transition near 866$~$nm (labelled IR for ``infrared''). 
A closed fluorescence cycle, which avoids pumping to dark metastable states, is established through the UV and IR lasers together with an additional magnetic field~\cite{barreto2023polarization}. The polarizations of the lasers are set to be linear and orthogonal to the magnetic field  which has a magnitude of $B\sim 4$~G. Then both lasers have $\sigma_+ + \sigma_-$ polarization in the atomic (magnetic) eigenbasis of the ion. In such a configuration, the atomic spectrum will show four dark resonances, as in Fig.~\ref{fig:fig1}(d).

We measure atomic spectra by collecting scattered ultraviolet light of the ion for different values of IR detuning $\Delta_{\text{IR}}$ while keeping the UV detuning $\Delta_{\text{UV}}$ fixed.
To ensure that the temperature of the ion is the same for all data points, we perform a pulsed algorithm where we interleave cooling and detection~\cite{barreto2022three,reiss2002raman}. A scheme of the algorithm is represented in Fig.~\ref{fig:fig1}(c). We start by performing a Doppler cooling stage for 100~$\mu$s with the UV laser red-detuned around $\Delta_{\text{UV}}=-15$~MHz and the IR laser blue-detuned around $\Delta_{\text{IR}}=+20$~MHz. With these cooling parameters, which we use for all measurements, we prepare the ion in a thermal state at a temperature close to the Doppler limit for a single well-compensated ion. After this, we switch $\Delta_{\mathrm{UV}}$ to $-25$~MHz and $\Delta_{\mathrm{IR}}$ to the measurement value and collect the scattered light of the ion for 50~$\mu$s. This time interval is chosen so that the scattering does not affect significantly the temperature of the ion. We repeat this sequence 20 times, and then perform a longer Doppler cooling stage of 1~ms to avoid excess heating of the ion and to remedy heating from other extreme events such as background gas collisions. Each sub-loop is repeated 100 times to reduce statistical uncertainties. Each datapoint then consists of 2000 measurements. 

In order to compensate excess micromotion for a single ion we use the DC endcaps, one compensation electrode and the atom oven, located below the trap. For the compensation we use the following algorithm. First, we set $\Delta_{\mathrm{IR}}=+20$~MHz and $\Delta_{\mathrm{UV}}\sim -\Omega_\text{RF}$, where $\Omega_\text{RF}$ is the drive frequency.
Then, we manually sweep the voltage of each compensation electrode and look for a minimum of the fluorescence. This reduces the power in the first-order modulation sideband of the UV spectrum, hence finding the position of the ion with minimum micromotion along the propagation direction of the UV laser~\cite{keller2015precise}. 
In order to do a full 3D compensation, this procedure should be iterated with three different lasers coming from from three different directions, ideally orthogonal between them, finding a global minimum. For the experiments presented in this paper, we only compensate the ion in two directions.

\section{Dark resonance spectra with micromotion}
\label{sec:sec3}
In this section we present measurements of dark resonance spectra including micromotion effects. We then introduce a theoretical model that fits the measurements and subsequently describe how to retrieve relevant experimental parameters with it.

\subsection{Experimental spectra}

\begin{figure*}[th]
\centering
\includegraphics[width=0.9\textwidth]{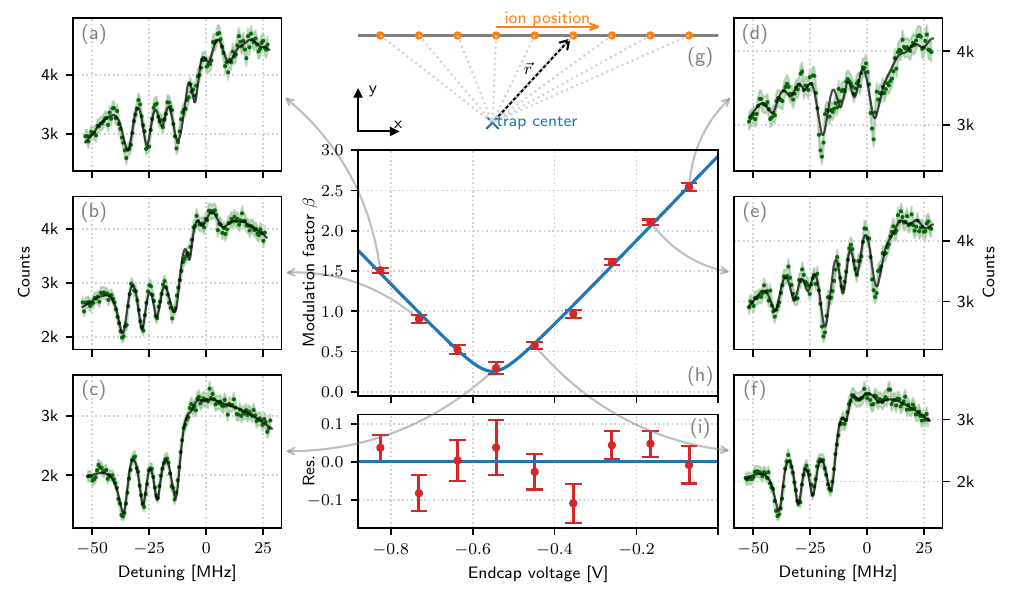}
\caption{\label{fig:fig2}
Dark resonance spectra with micromotion induced by (de)compensation of the position of the ion by means of an endcap voltage. Subfigures (a)-(f) depict the acquired spectra (dark-green dots) and the corresponding fitted model (black lines) for selected voltage values. The parameter $\beta$ of the regression model serves as a measure of micromotion modulation and is expected to be proportional to the distance from the trap center. The ion position for each measurement varies proportionally to the applied voltage on the endcap, along a linear trajectory as illustrated in (g). The values of the factor $\beta$ as a function of the applied voltage are plotted in (h) as red dots, accompanied by the expected hyperbolic relation fitted to the data (blue line). Residuals of the fit are presented in~(i).
}
\end{figure*}

We measure the dark resonance spectra of a single ion for different positions in the ion trap where it is subjected to varying degrees of micromotion. As we vary the potential applied to the top endcap electrode, $V_{\mathrm{ec}}$, the ion is displaced to different positions with respect to the center of the trap following a straight line, as shown at the sketch in Fig.~\ref{fig:fig2}(g).
When the ion is closer to the center we expect it to experience less micromotion than when it is further away. For each position a spectrum was measured and the modulation factor was determined by a fit of the model.
Some representative spectra are shown in Figs.~\ref{fig:fig2}(a)-(f). The one with the least micromotion, when the ion is closest to the center of the trap, corresponds to Fig. \ref{fig:fig2}(c). There, the four fundamental dark resonances can be identified. Further away from the center, when micromotion is more pronounced, echoes of the four fundamental resonances appear in the spectrum, as can be seen in the other subplots.  

For all spectra in Fig.~\ref{fig:fig2}, the data are plotted along with fits to a model which considers micromotion. In the following subsection we show how to build such a formalism, which adds a modulation at the drive frequency $\Omega_\text{RF}$ in the atomic spectra. This allows us to retrieve a modulation factor that quantifies the amount of micromotion in the ion, as well as the temperature of the ion.

\subsection{Numerical calculation of spectra including micromotion}

To calculate the spectra including micromotion effects we closely follow~\cite{barreto2022three} and~\cite{oberst1999resonance}, in which similar theoretical approaches were developed. To model the atomic dynamics of the full eight-level system interacting with the two lasers, we resort to a master equation of the form
\begin{equation}
    \frac{d\rho}{dt} = - \frac{i} {\hbar} [H,\rho] + \mathcal{L}(\rho)\,.
\end{equation}
Here, $\rho$ is the density matrix of the system, $H$ is the Hamiltonian of the eight-level system interacting with the two lasers under the rotating wave approximation in the frame rotating with the lasers, and $\mathcal{L}$ is a Lindblad superoperator that models the spontaneous emission from the upper levels into the two lower manifolds, as well as finite laser linewidths and thermal motion as described in~\cite{rossnagel2015fast}. 

To include micromotion in the dynamics, we consider the Doppler shift that the ion experiences due to its periodic motion with respect to the beams. This can be treated by adding the corresponding Doppler shift to the detuning of each laser for the ion at rest, $\Delta_{i_0}$, as
\begin{equation}
    \Delta_i = \Delta_{i_0} + \vec{k}_i \cdot \vec{v}_0 \cos(\Omega_{\text{RF}} t), 
\end{equation}
where $i=\{$IR,UV$\}$, $\vec{k}_i$ is the wave vector of the corresponding laser, $\Omega_\text{RF}$ is the drive frequency and $\vec{v}_0 \cos(\Omega_{\text{RF}} t)$ is the instantaneous velocity of the atom, considering only the contribution due to micromotion. The magnitude of $\vec{v}_0$ is proportional to the displacement of the ion from the trap center. Motion due to thermal effects is introduced in the following section.

We define a dimensionless modulation factor $\beta$ for each beam as
\begin{equation}
    \beta_i\Omega_\text{RF} = \vec{k}_i \cdot \vec{v_0}\,,
\end{equation}
which quantifies the strength of micromotion effects~\cite{berkeland1998minimization}. It is important to note that the value of $\beta_i$ depends on the product between the wavevector of the corresponding beam and the driven ion motion. 
As in our case both beams are co-propagating, we have that ${866}\,\text{nm}\times\beta_{\text{IR}} = 397\,\text{nm}\times\beta_\text{UV}$. Therefore only one of the $\beta_i$ needs to be specified as a free parameter in the mathematical model. 
For the sake of simplicity we refer to $\beta \equiv \beta_{\text{UV}}$ and use it as the free parameter that quantifies the micromotion.

The explicit time-dependence of the dynamics is treated through a Floquet-like approach, where we map the master equation to a Liouvillian equation of the form
\begin{equation}
    \frac{d\vec{\rho}}{dt} = L(t)\vec{\rho},
    \label{eq:lio}
\end{equation}
where $\vec{\rho}$ is the vectorized form of $\rho$ and $L(t)$ is a Liouvillian superoperator in the corresponding matrix form. This can in turn be written as
\begin{equation}
L(t)=L_0+\beta\,\Omega_{\text{RF}}\,\Delta L\,\cos(\Omega_{\text{RF}} t),
\end{equation}
where $L_0$ is the Liouvillian for the system without modulation, and $\Delta L$ is a $8^2 \times 8^2$ matrix with constant elements~\cite{oberst1999resonance}. 

For the trivial case where $\beta=0$, i.e. without micromotion, the steady state can be found by inverting $L_0$ with a normalization constraint over $\vec{\rho}$. We calculate the atomic spectrum by plotting, as a function of the detuning $\Delta_{\text{IR}}$, the sum of the populations of the two excited P states $\rho_{_{\text{P}_{+1/2}\text{P}_{+1/2}}}+\rho_{_{\text{P}_{-1/2}\text{P}_{-1/2}}}$, which is proportional to the fluorescence of the ion. The plot of Fig.~\ref{fig:fig1}(d) was calculated in this way, and exhibits the four expected dark resonances.

If $\beta \neq 0$, the atomic populations will oscillate around a mean value with frequency components that are multiples of $\Omega_\text{RF}$. This will be reflected in the spectrum as modulation sidebands displaced from the original dips by integer multiples of $\Omega_\text{RF}$. To obtain the time-averaged value of the populations, we propose a solution for times longer than all transient times of the system~\cite{barreto2023transient} with the form
\begin{equation}
    \vec{\rho}(t)=\sum_{m=-\infty}^{+\infty} \vec{\rho}_m e^{im\Omega_\text{RF} t}.
    \label{eq:sum}
\end{equation}
When introducing this ansatz into Eq.~\eqref{eq:lio}, we get the recursive relation
\begin{equation}
    (L_0 + in\Omega_\text{RF})\vec{\rho}_n + \frac{\beta\Omega_\text{RF}}{2}\Delta L (\vec{\rho}_{n+1}+\vec{\rho}_{n-1}) = 0.
    \label{eq:recursive}
\end{equation}

We are interested in obtaining $\vec{\rho}_0$, i.e. the mean value of $\vec{\rho}(t)$. To do so, we define two operators of the form
\begin{equation}
    S^{+}_{n-1} = -\left(L_0 -in\Omega_\text{RF} + \frac{\beta\Omega_\text{RF}}{2}\,\Delta L\, S_n^{+}\right)^{-1}\frac{\beta \Omega_\text{RF}}{2}\Delta L,
    \label{Splus}
\end{equation}
\begin{equation}
    S^{-}_{n+1} = -\left(L_0 -in\Omega_\text{RF} + \frac{\beta \Omega_\text{RF}}{2} \, \Delta L\, S_n^{-}\right)^{-1}\frac{\beta \Omega_\text{RF}}{2}\Delta L.
    \label{Sminus}
\end{equation}
It is straightforward to see that $\vec{\rho}_{n+1} = S_n^{+}\vec{\rho}_n$ for $n \geq 0$ and $\vec{\rho}_{n-1}=S_n^{-}\vec{\rho}_n$ for $n \leq 0$, so the defined operators act as raising and lowering operators respectively, in the sense that they relate components that oscillate at neighbouring frequencies. 

In terms of these operators, we can evaluate Eq.~\eqref{eq:recursive} for $n=0$ and obtain
\begin{equation}
    \left( L_0 + \frac{\beta\Omega_\text{RF}}{2}\,\Delta L \,(S_0^+ + S_0^-) \right)\vec{\rho}_0 = 0.
\end{equation}
Finally, to calculate $S_0^{+/-}$ we truncate the sum in Eq.~\eqref{eq:sum} such that $\vec{\rho}_{\pm N_{max}}=0$. This is equivalent to saying that we will include only up to $N_{max}-1$ modulation sidebands in the spectrum. The appropriate value of $N_{max}$ depends on the value of $\beta$: for larger micromotion, 
a higher number of sidebands will appear, and a higher $N_{max}$ will be required to properly describe the atomic spectra. One can choose $N_{max}$ by imposing that the Bessel function $J_{N}(\beta)$ is negligible for $N>N_{max}$. Nevertheless, setting $N_{max} \sim \beta$ is sufficient for all practical considerations. For all fits in this work, we set $N_{max}=5$, since $\beta$ never surpassed that value in our experiments.

\subsection{Estimation of the modulation parameter}
In Figs.~\ref{fig:fig2}(a)-(f) we plot, in solid line, a fit to the model for the six measured spectra, which are shown in green dots along with their uncertainties. The fits are in very good agreement with the measurements, evidencing the robustness of the model to predict how micromotion affects the atomic spectra. From them, we retrieve a value for the modulation factor $\beta$ for each voltage, as well as the temperature $T$ of the ion as explained in the next section. 

In Fig. \ref{fig:fig2}(h) we show how $\beta$ depends on the endcap voltage $V_{\mathrm{ec}}$. There is an optimal point at $-0.55(1)~$V in which $\beta$ is minimum, but not zero. This is due to extra stray fields not compensated in directions that are orthogonal to the one of the endcaps.

Considering that the ion moves along a trajectory that misses the center of the trap while sweeping $V_{\mathrm{ec}}$ by an amount of $y$, that its position varies linearly with it, as it is shown in 
Fig.~\ref{fig:fig2}(g), following $(V_{\mathrm{ec}}-V_0)\alpha_V$
and the proportionality of $\beta$ with the distance $r$ to the center ($\beta=\alpha_\beta\,r$),
the data can be modeled by a hyperbola function of the form
\begin{equation}
    \label{eq:hiperbola}
    \beta(V) = \alpha_\beta\,\sqrt{\alpha_V^2\,(V-V_0)^2+y^2}
    \;\;\;.
\end{equation}
Bringing the $\alpha_\beta$ inside the square root, the four parameters in the equation above can be grouped and reduced to three in the regression model. The fitted model is plotted in a solid blue line over the experimental data in Fig.~\ref{fig:fig2}(h), showing good agreement with the data. From this fit we extract the minimum value of $\beta$ which is $\beta_{min}=0.30(5)$. Furthermore, in Fig.~\ref{fig:fig2}(i) we plot the residuals of the fit, showing a random dispersion around zero.

\section{Micromotion and thermal effects}
\label{sec:sec4}

\subsection{Inclusion of thermal effects in the spectrum}
We incorporate the temperature $T$ of the ion in the model following the approximation presented in~\cite{rossnagel2015fast}, where the effect is accounted for as an additional dephasing term proportional to the square root of $T$, which is added as an extra broadening of the laser linewidths
\begin{equation}
    \Gamma_D = \left| \vec{k}_{\text{IR}} - \vec{k}_{\text{UV}} \right| \sqrt{\frac{k_B T}{2m}}.
    \label{ec:dopplerbroadening}
\end{equation}
One can understand this approximation the following way. The Doppler effect generates a shift proportional to $k v$. As we are analyzing a two photon transition, we replace $k$ for the difference $\left| \vec{k}_\text{UV}-\vec{k}_\text{IR}\right|$. Because we are considering a thermal state, $v$ fluctuates with a standard deviation $\sqrt{k_B T/m}$. This is modelled through stochastic shifts in the frequency of the lasers.

This method implies then two important approximations, the first by only considering two-photon processes, and the second by approximating a Gaussian distribution by a Lorentzian one. The $\sqrt{2}$ in the denominator of Eq.~(\ref{ec:dopplerbroadening}) was chosen because it was found to provide the best approximation. We note, however, that the proportionality factor leading to the best results is generally dependent on the temperature regime of interest as well as on the relevant linewidths. A more detailed analysis of this is beyond the scope of this paper and will the subject of a future publication. Finally, to include $\Gamma_D$ in the model, we multiply it by $k_i/(k_\text{UV} + k_\text{IR})$, where $i = \{\text{UV}, \text{IR}$\}, and add it to the corresponding linewidth of the laser.

\begin{figure}[h]
\centering
\includegraphics[width=0.5\textwidth]{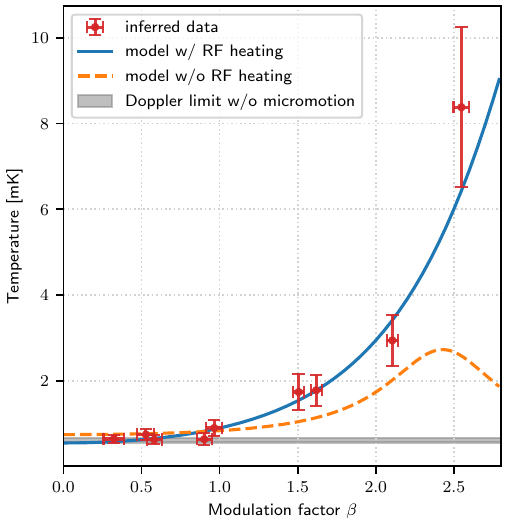}
\caption{\label{fig:fig3} 
Relationship between temperature and micromotion modulation factor $\beta$. The data inferred  from the spectra in Fig.~\ref{fig:fig2} are depicted by red dots along with uncertainty intervals (derived from the standard error of the fitted parameters in Fig.~\ref{fig:fig2}). Two models have been tested to describe this relationship: one excluding RF heating (orange dashed line) and another incorporating it (blue straight line). Further details about each model can be found in the main text. The estimated temperatures for low micromotion are consistent with the Doppler limit for detuning of around -10 MHz, represented by the gray shaded area in the graph.
}
\end{figure}

Following this procedure, the ion temperature $T$ becomes a parameter that we can retrieve from the fits. The temperatures estimated from the spectra of Fig. \ref{fig:fig2} are plotted as a function of the modulation factor $\beta$ in Fig.~\ref{fig:fig3}. The plot shows a temperature that is approximately constant and close to the Doppler limit for $\beta < 1$. For larger micromotion, the temperature exhibits a significant boost upon increasing $\beta$. To understand this behavior, we now characterize the RF heating that the ion experiences and add this effect to the equation describing the thermalization of the ion.

\subsection{Thermalization in the presence of micromotion}
When the ion is close to the Doppler temperature, the equation determining the energy balance under laser cooling is
\begin{equation}
\dot{E} = \dot{E}_\mathrm{Dop} + \dot{E}_\mathrm{rec} + \dot{E}_\mathrm{RF},
\end{equation}
where $\dot E_{\rm Dop}$ represents the Doppler cooling mechanism due to the interaction of the atom with the lasers, $\dot E_{\rm rec}$ is the recoil heating due to the random emission of one photon of the atom, and $\dot E_\mathrm{RF}$ is the RF heating. In the following, we show how to calculate the three terms separately.

\subsubsection{Doppler cooling}
Doppler cooling works as a dissipation channel~\cite{yan2019analytical}. It acts on the ion through a force of the form
\begin{equation}
    \vec{F}_d = \hbar\vec{k}\Gamma \rho_{ee}(\Delta,\beta),
\end{equation}
where $\vec{k}$ is the wave vector of the Doppler cooling laser, $\Gamma$ is the linewidth of the cooling transition, and $\rho_{ee}$ is the excited state population, that is modulated by the micromotion. We will approximate the dynamics by a two-level system corresponding to the S$-$P transition which is the most relevant for the Doppler cooling. Thus, all laser parameters in this subsection will correspond to the UV laser.

In the low saturation limit we can express the excited state population using Bessel functions $J_n$ as~\cite{devoe1989role}
\begin{equation}
    \rho_{ee}(\Delta,\beta)=\left( \frac{\mu E_0}{2\hbar} \right)^2 \sum_n \frac{J_n^2(\beta)}{(\Delta+n\Omega_\text{RF})^2 + (\Gamma/2)^2},
\end{equation}
where $\mu$ is the electric dipole moment of the transition and $E_0$ is the electric field of the laser.
To include the motion of the atom, we will consider low velocities $\vec{v}$ such that  $|\vec{k}\cdot \vec{v}| \ll |\Delta|$, where $\vec{k}$ is the wave vector of the cooling laser. We can then approximate
\begin{equation}
\rho_{ee}(\Delta+\vec{k}\cdot \vec{v},\beta) \sim \rho_{ee}(\Delta,\beta) + \frac{d\rho_{ee}(\Delta,\beta)}{d\Delta}(\vec{k}\cdot\vec{v}).
\end{equation}

Considering that $\langle \vec{v} \rangle = 0$ since the atom is trapped in a harmonic potential, the cooling power turns to be
\begin{equation}
    \dot{E}_\mathrm{Dop} = \langle\vec{F}_d \cdot \vec{v} \rangle = - \hbar |\vec{k}|^2 \Gamma \frac{d\rho_{ee}(\Delta,\beta)}{d\Delta} \langle v_{\hat{k}}^2 \rangle,
\end{equation}
where $v_{\hat{k}}$ is the velocity of the ion projected in the direction of $\vec{k}$. Finally, assuming a thermal state we relate the mean value of the squared of the velocity with the temperature as $\frac{1}{2}m\langle v_{\hat{k}}^2 \rangle=\frac{1}{2}k_BT$. In this form, we can write $\dot{E}_\mathrm{Dop}$ in terms of the temperature of the ion.

\subsubsection{Recoil heating}
Recoil heating is a mechanism that balances Doppler cooling leading to a non-vanishing temperature limit. It is due to the spontaneous emission of a photon when the atom is excited, which causes a recoil kick of the atom in a random direction. The rate of heating is proportional to the energy of the emitted photon, the decay rate $\Gamma$ and the emission probability linked to the excited state population $\rho_{ee}$, resulting in
\begin{equation}
    \dot{E}_\mathrm{rec}=\frac{(\hbar |\vec{k}|)^2}{2m}\Gamma \rho_{ee}(\Delta,\beta).
\end{equation}

\subsubsection{RF heating}
The RF heating mechanism is given by electric noise at frequency components of the driving field in $\Omega_\text{RF} \pm \omega$, where $\omega$ is a secular frequency of the ion. 
This causes a periodic ponderomotive noise force at $\omega$ that adds extra heating. In~\cite{kalincev2021motional}, the resulting RF heating rate was shown to scale as
\begin{equation}
    \dot{E}_\mathrm{RF} \sim \left[ \vec{\nabla}\left(\vec E^2(\vec{r}) \right)\right]^2 \sim \beta^2\,.
\end{equation}
Here, we are using that the electric field $\vec E$ is approximately linear in the ion displacement from the trap center, and that the modulation factor $\beta$ is also proportional to this displacement. 
We can then express the rate for RF heating as
\begin{equation}
    \dot{E}_\mathrm{RF} = C_\mathrm{RF}\beta^2,
\end{equation}
where $C_\mathrm{RF}$ is a factor that depends on experimental parameters like the amount of electric noise. Therefore, it is a quantity that must be estimated from the measurements.

\subsubsection{Stationary temperature}
Considering all the above contributions one can find the stationary value of $T$ by setting $\dot{E}=0$. The expression for the temperature then becomes
\begin{equation}
    T_0 = \,\frac{\hbar}{k_B}\frac{\rho_{ee}(\beta,\Delta)}{\dfrac{\partial\rho_{ee}}{\partial\Delta}(\beta,\Delta)}+\frac{C_\mathrm{RF}}{\left( \dfrac{\hbar |\vec{k}|^2 \Gamma k_B}{m} \right)}\,\frac{\beta^2}{\dfrac{\partial\rho_{ee}}{\partial\Delta}(\beta,\Delta)}.
    \label{eq:finalexpression}
\end{equation}
If we set $\beta=0$ we obtain $T_0 = \frac{\hbar\Gamma}{2k_B}$, which corresponds to the well-known Doppler limit temperature. 

In Fig.~\ref{fig:fig3} we show a fit to Eq.~\eqref{eq:finalexpression} in solid blue line, leaving $C_\mathrm{RF}$ and $\Delta$ as fitting parameters. For comparison, we also include in orange dashed line a fit to the model without RF heating, i.e. setting $C_\mathrm{RF}=0$ and leaving $\Delta$ as the only fitting parameter. From this last fit, we obtain $\Delta = 2\pi \times -5.1(8)$~MHz. However, this model does not properly fit the measurements for $\beta>1$. In particular, neglecting RF heating leads to the prediction of a temperature decrease when $\beta$ becomes larger than $\sim2.4$, in clear contradiction with the experiments.

We can see that the inclusion of the RF heating term captures the increasing behavior of $T$ with increasing $\beta$, in good agreement with the measurements. From this fit we retrieve $\Delta = 2\pi \times (-11.7)(17)$~MHz, consistent with the detuning used to Doppler cool the ion before the measurements. 

On the other hand, we obtain $C_\mathrm{RF}=8.4(32)\times 10^{-20}$~J/s. We can rewrite it in terms of a phonon heating rate of one motional mode of the ion~\cite{leibfried2003quantum}. Considering one of the radial modes of the ion, with $\omega=2\pi \times 1~$MHz, the heating rate becomes
\begin{equation}
\dot{n}_\mathrm{RF}=0.20(7)\times \beta^2~\mathrm{ph/s}. 
\end{equation}
This result is in good agreement with other similar measurements of RF heating effects~\cite{keller2015precise}.

\section{Multi-ion spectra}
\label{sec:sec5}

In a final phase of this work, we measure spectra for a system of two ions. The setup used collects the light of all the trapped ions globally on a photo-multiplier tube, not allowing to distinguish each one individually, but allowing fast measurements with low background. The spectra are therefore composed of the light emitted by all ions. As each ion is in a different position in the trap, they will experience different micromotion, and therefore contribute differently to the summed spectrum. We aimed at extracting the modulation factor and temperature of each ion from the global signal.

\begin{figure}[h]
\centering
\includegraphics[width=0.5\textwidth]{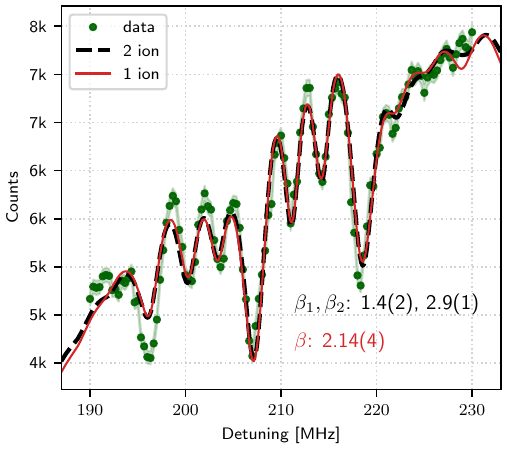}
\caption{\label{fig:twoions} 
Dark resonance spectrum acquired by collecting light from two trapped ions. The data are compared with the fitted model for one (red line) and two (dashed-black line) ions. The $\beta$ obtained in each regression are reported in their respective colors.
}
\end{figure}

In Fig. \ref{fig:twoions} we show one typical spectrum for two ions and different model fits. We find that a two-ion model, with an independent modulation factor $\beta$ for each one, adequately fits the data, but the retrieved parameters are very sensitive in their initial guess. Moreover, we see that a one-ion model, with only one $\beta$, can also produce reasonable fits. Also, we studied the behaviour of the fitted modulation factors as a function of the position of the ion crystal in the trap, and found that the results obtained were also very sensitive to the initialization of the parameters. 

We conclude that extracting independent modulation parameters or temperatures for multiple ions from a single signal is an ill-defined problem, with too many free parameters. For an appropriate measurement, independent collection of each ion's fluorescence would be necessary. This has to be combined with the short collection times (50~$\mu$s), which ensures that the temperatures of the ions are marginally affected by the measurement. Such a procedure is rather challenging with electron multiplying charge-coupled devices (EMCCDs) or complementary metal–oxide–semiconductor (CMOS) cameras, but multi-pixel photomultiplier tubes, avalanche photodiode detectors or photon counting cameras could provide useful alternatives. 

\section{Conclusions} \label{sec:conclusions}

We have analyzed the effects of micromotion in dark-resonance atomic spectra. We presented experimental results obtained with a single trapped calcium ion for which the amplitude of the RF-driven motion was varied applying a controllable voltage to one trap electrode. We introduced a theoretical model that accurately describes the corresponding spectra, and used it to fit our measurements. From these fits, we were able to retrieve the micromotion modulation factor of each curve.

The same fits were used to estimate the temperature of the ion under different micromotion amplitudes. Our results evidence the presence of a micromotion-induced heating mechanism, also known as RF heating. We  include this phenomenon in the standard description of Doppler cooling and show that it leads to good predictions of the final temperature of the ion depending on the value of the micromotion modulation parameter. 

These methods could be used to study heat transport in trapped ion systems where excess micromotion is unavoidable due to crystal geometry, for example 2D crystals in linear traps. However, we note that a proper description of the dynamics of several ions requires single-ion detection. Indeed, our studies of the spectra obtained by collecting the light emitted by two ions together indicate that global light collection is not sufficient for a good characterization of the motional state.

\medskip

\section*{Data availability statement}
The original contributions presented in the study are included in the article; further inquiries can be directed to the corresponding author.

\section*{Author contributions}
CTS and NANB were responsible for planning the research objectives an goals. NANB, ML and MB were responsible for carrying out the experiments and for the analysis of the measured data. NANB developed the numerical models together with MB and CC. The theoretical methods were supervised by CTS and CC. ML was in charge of producing the final figures. NANB was in charge of writing the article, with contributions from all authors who approved the submitted version.

\section*{Funding}
NANB, MB, ML and CTS acknowledge support for grants PICT2018-03350, PICT2019-04349 and PICT2021-I-A-01288 from ANPCyT (Argentina) and grant UBACYT 2018 Mod I - 20020170100616BA from Universidad de Buenos Aires.
CC acknowledges funding from grant PICT 2020-SERIEA-00959 from ANPCyT (Argentina).

\section*{Acknowledgements}
We acknowledge the contribution of M. Drechsler in the building of the ion trap and early experiments. 
We thank F. Schmidt-Kaler for his invaluable support and generosity, as well as J. P. Paz for his unconditional help in the setting up of the laboratory.
Finally, we thank F. Meconi for taking the photo of Fig.~\ref{fig:fig1}(a).

\section*{Conflict of interest}
The authors declare that the research was conducted in the absence of any commercial or financial relationships that could be construed as a potential conflict of interest.

\bibliography{bibpaper}

\end{document}